\documentclass[myclassdoc,rjpdebug]{rjparticle}
\usepackage{graphicx}
\usepackage{colortbl,multirow, hhline}
\newcommand{\graycline}{\arrayrulecolor{gray}\cline{2-8}\arrayrulecolor{black}}

\title{Modeling Alpha Particle-Induced Radioluminescence Using Geant4} 

\author[1,$a$]{C. OLARU}

\author[1,$b$]{M.-R. IOAN}

\author[1,$c$]{M. ZADEHRAFI}

\affil[1]{``Horia Hulubei'' National R\&D Institute for Physics and Nuclear Engineering,\\
Reactorului 30, RO-077125, P.O.B. MG-6, M\u{a}gurele-Bucharest, Romania, EU\\
\emailca{b}{razvan.ioan@nipne.ro}\\
\email[a]{claudia.olaru@nipne.ro},
\email[c]{mastaneh.zadehrafi@nipne.ro}}


\begin{document}
\maketitle
\begin{abstract}
Optical detection of alpha particle emitters in the environment by air radioluminescence is a new technology that enables sensing a radiological threat at safe distances, without putting personnel at risk or contaminating equipment. Radioluminescence detection systems need to be fine-tuned to efficiently capture a substantial number of photons while minimizing the contribution from ambient ultraviolet light. The accurate simulation of radioluminescence, in conjunction with ray tracing, facilitates the design and optimization of such detection systems. In this work, an application within the Geant4 framework has been developed to simulate radioluminescence photons emitted in the vicinity of accelerated alpha particles and at the surface of alpha radioactive samples. The application relies on existing scintillation physics implemented in Geant4 classes such as G4OpticalPhysics and G4Scintillation, which are used to simulate radioluminescence photons as scintillations produced during the passage of alpha particles through air. The application computes the ultraviolet image of alpha particles accelerated at energies of 5.1 MeV and 8.3 MeV, as well as an extended alpha source \cite{1Luchkov}. The application enables optimization of experimental setups for various scenarios, such as radiological emergency management, radiological crime scene investigations, or decommissioning of nuclear facilities, thus minimizing the use of costly resources and exposure to radiation.
\end{abstract}

\section{Introduction}
Alpha radiation detection poses a real challenge to radiological emergency response teams. Due to the short range of alpha particles in air (about 4 cm at 5 MeV), conventional detectors for alpha contamination (\eg, silver-activated ZnS thin films, passivated implanted planar silicon, and silicon gold surface-barrier detectors) usually rely on direct interactions between alpha particles and the sensitive detector material \cite{2Vajda}. This approach requires scanning near contaminated surfaces (\ie, within the range of alpha particles in air), which complicates the management of emergencies and decontamination efforts by exposing operators to hazards and health risks (\eg, fire, radiation, toxic substances, etc.) and risking contaminating the detectors. Moreover, the use of short-range handheld detectors makes scanning large areas and complex terrain geometries (\eg, collapsed buildings) laborious and time-consuming.

The shortcomings of conventional detectors can be overcome by applying a detection technique based on the alpha particle-induced ultraviolet (UV) luminescence of air, the so-called radioluminescence \cite{3Huggins}. In this approach, the atmosphere serves as a scintillator. The ionization of the atmosphere triggered by the passage of alpha particles creates a cloud of excited air molecules which decay radiatively by emitting UV photons. Most of the emission is due to the de-excitation of molecular nitrogen (N$_2$), and to a much lesser extent, due to trace amounts of nitric oxide (NO) in the composition of air, spanning three ultraviolet spectral regions. About 99\% of the emissions occur in the UV-A and UV-B spectral regions, from 280 nm to 440 nm \cite{4SandThes}. The UV light can propagate through air several hundreds of meters, many orders of magnitude larger than the range of alpha particles (primary radiation) in air \cite{5Goody}, enabling sensing a radiological threat at a safe distance.

An airborne application of radioluminescence detection technology can support emergency preparedness and management in case of accidental or intentional dispersal of alpha-emitting radionuclides in the environment \cite{6Schneider}. Novel optical technologies for such applications, including calibration systems and methodologies, have been developed in the framework of the European Metrology Programme for Innovation and Research (EMPIR) project 19ENV02 RemoteALPHA \cite{7RemoteA}. Instrumentation and procedures developed in this project can also support national and international authorities in preventing the illicit trafficking of alpha-emitting materials. Furthermore, this instrumentation will benefit the nuclear industry by providing a detection system able to remotely monitor the manufacturing, handling, and storage of alpha-emitting materials.

To facilitate emergency management, these detection systems should be optimized by maximizing the radioluminescence throughput. This is done by using large receiving optics, low noise photomultipliers (PMT), and suppressing the background signal through efficient UV band filtering. At the same time, the systems need to have an optimal field of view to allow mapping alpha contaminations when mounted on tripods or UAVs. In the framework of RemoteALPHA, the systems have been characterized and optimized using the joint metrological infrastructure of participating institutions. 

A useful alternative for prototyping detector instrumentation is Monte Carlo modeling \cite{8MetroMC,9DoseDist,10RPP}. With accurate models of involved physical processes, these development tools allow reducing the experimental component to a minimum making the prototyping more accessible in terms of required expenditure, equipment, and services. The Geant4 toolkit is well-suited for simulating the radioluminescence generation, which can be achieved by generating optical photons in well-defined scintillating materials. The performance of the MC simulation toolkit was tested by modeling the generation of air scintillation (luminescence) induced by the ionizing effects of alpha sources. The ultraviolet image of the alpha source can be obtained by registering the spatial distribution of the scintillation photons generated within a luminescent medium.

\section{Simulation method}
When alpha particles in the megaelectronvolt energy range propagate through air, they ionize various molecules and atoms. The released electrons, the so-called secondary or delta electrons, also interact with the air molecules – mainly with N$_2$ – and generate further low-energy electrons. This cascade-like process leads to the excitation of air molecules, which then relax to lower energy electronic states by emitting luminescent photons in the UV spectral region \cite{8Arqueros}. Secondary electrons can be very energetic, up to a few kiloelectronvolt, but those with energies in the range of 14 eV to 15 eV contribute the most to triggering radioluminescence production, according to available maximal excitation cross-section values \cite{9Itikawa}.

On average, a single 5 MeV alpha particle generates in air about 100 photons in the UV-A and UV-B spectral regions \cite{10Sand}. Spanning these regions, the radioluminescence emission is given by the second positive system - 2P ($C^3\Pi_u\rightarrow B^3\Pi_g$) and the first negative system - 1N ($B^2\Sigma_u^+\rightarrow X^2\Sigma_g^+$) of the nitrogen molecule. In contrast, the emission yield between 200 and 280 nm (UV-C) is considerably lower, at about 0.9 photons per MeV of deposited energy \cite{4SandThes}. However, this yield can be boosted by more than three orders of magnitude, by adding trace amounts of nitric oxide to the nitrogen atmosphere surrounding an alpha source \cite{11Kerst}. NO produces radioluminescence over the $\upgamma$ band system ($A^2\Sigma^+\rightarrow\ X^2\Pi$), between 200 nm and 300 nm, which is almost exclusively within the UV-C spectral region. This band system is mainly produced by excitation transfer from N$_2$ to NO molecules \cite{12Kerst}. The radiative transitions between the vibrational levels (denoted as $\nu$) of these specific energy states shape the radioluminescence spectrum used in the remote detection of alpha sources in air \cite{1Luchkov}.

Studies on modeling the generation of radioluminescence photons from radioactive sources using Geant4 are relatively scarce. Roberts \cite{13Roberts} simulated the solar blind photon flux and 2D images of alpha, beta, and gamma sources by recording the radioluminescence photons produced at 10 m from the radioactive source on a 40 cm diameter detector. Although the simulation method used was not described in detail, it provides valuable information related to radioluminescence modeling using the Geant4 toolkit. Another work is the one of Thompson \etal \cite{14Thompson} who, based on molecular ionization, excitation, and emission models, developed a new physics model for predicting the number of photons emitted in the UV-A and UV-B regions from excited states of atmospheric nitrogen. The “Air Fluorescence Model” \cite{14Thompson} was implemented in the Geant4 framework and used for studying the UV photons induced in air by alpha, gamma and beta radiation emitted by radioactive sources, as well as investigating the localization of Am-241 ($\upalpha$-emitter), Co-60 ($\upgamma$-emitter) and P-32 ($\upbeta$-emitter), through air fluorescence (radioluminescence). The model was also used for predicting the fluorescence yields for Po-210, Am-241, U-235, Co-60, P-32 and Sr-90.

This work is focused on providing a simplified approach to modeling alpha induced radioluminescence production in Geant4. The simulation was performed using integrated physics models, rather than implementing a dedicated one for simulating the atomic physics processes involved in air luminescence. This way, the performance of Geant4 capabilities in modeling this effect was studied by employing the OpNovice pre-existing example. Building on this example, the simulation was expanded to consider air in normal atmospheric conditions (1 atm pressure and 22°C temperature) as a scintillating material. The primary particles are set to alpha particles, and the generated radioluminescence comes in the form of secondary radiation as optical photons, if the processes from the G4Scintillation source class are enabled. The scintillating medium is characterized by its optical properties such as photon emission spectrum, scintillation yield, scintillation time constants, etc. To model alpha particles emitted by radioactive sources, such as a structured Am-241 source and a beam of accelerated alpha particles, the G4ParticleGun source class is used. In both cases, the default electromagnetic physics processes were replaced with the ones from the G4EmLivermorePhysics model class to enable the generation of low energy (minimum 10 eV) electron interactions with high accuracy below 100 keV. Lastly, the generation of optical photons from excited air molecules is activated by instancing the G4OpticalPhysics source class in the main simulation source code, OpNovice.cc. By adding it to the list of physics constructors together with G4EmLivermorePhysics and G4RadioactiveDecayPhysics source classes, all physics interactions leading to radioluminescence production are enabled. Fig. 1 shows schematically the relevant source classes used in the simulation of alpha particle-induced radioluminescence.

\begin{figure}
    \centering
    \includegraphics[width=0.8\linewidth]{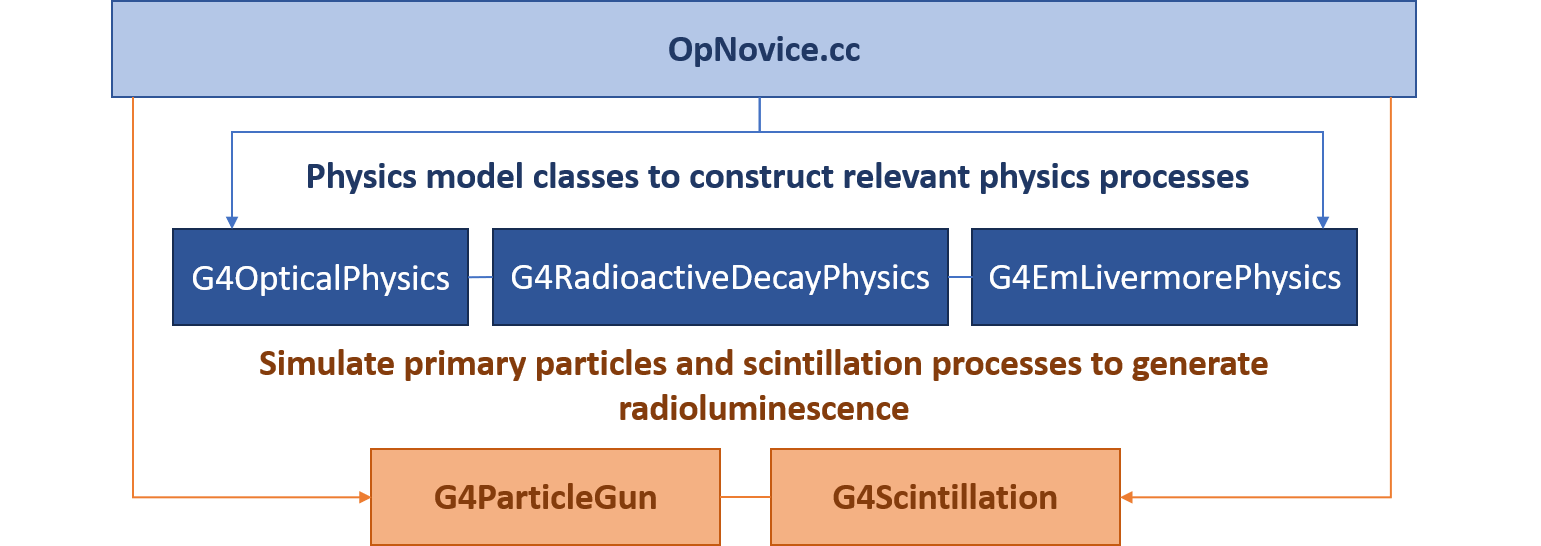}
    \caption{Schematic representation of Geant4 source classes used in radioluminescence modeling.}
    \label{pic1}
\end{figure}

The following method describes how to simulate the generation of radioluminescence using semi-empirical radiative transition parameters of the N$_2$ and NO molecules. The optical properties of a scintillating material are provided by the user and stored as vector entries in a material property table (MPT). The MPT is instanced in the detector description user class OpNoviceDetectorDescription and linked to the scintillation processes. To use air as the scintillating medium, the radiative transition parameters shown in Table 1 were registered as optical properties in the MPT. The most notable parameters used in the simulation are: the band origin of the transition (wavelength), the Franck-Condon factor - $q_{\nu'\nu''}$ (transition probability), the Einstein coefficient - $A_{\nu'\nu''}$ (spontaneous emission rate), the radiative lifetime - $\tau_{\nu'}$, and the branching ratio $B_{\nu'\nu''}$, calculated using (1), where $\nu'$ denotes the vibrational level of the higher energy state, $\nu''$ is the vibrational level of the lower energy or ground state of a molecule, and \(\nu = 0, 1, 2, ...\).

\begin{equation}\label{eq1}\begin{aligned}
B_{\nu'\nu''}=\frac{A_{\nu'\nu''}}{\sum_{\nu''}A_{\nu'\nu''}}
\end{aligned}\end{equation}\\

In the MPT both energy dependent and constant optical properties can be registered. Some of the properties used in the simulation are displayed in Table 1, where the 2P, 1N and $\upgamma$ band systems (BS) are marked out in column 0. Then, column 1 depicts the wavelengths of the identified spectral lines from radioluminescence measurements performed at the PTB Ion Accelerator Facility (PIAF) \cite{1Luchkov}, while column 2 contains the band origins available in the scientific literature, characteristic to the $\nu'\rightarrow\nu''$ vibronic transitions, listed in Table 9 and Table 15 of reference \cite{15Gilmore} for the 2P and 1N systems, respectively, and in Table III of reference \cite{16Luque} for the $\upgamma$ band system. The values displayed in the following columns correspond to these wavelengths. Columns 3 and 4 display the vibrational levels of the upper energy ($\nu'$) and lower energy ($\nu''$) electronic states involved in each radiative transition. Columns 5 and 6 contain the values of the Franck-Condon factors and Einstein coefficients, respectively, extracted from the previously mentioned tables of references \cite{15Gilmore} for N$_2$ and \cite{16Luque} for NO. Column 7 is filled with branching ratios calculated using (1). Lastly, column 8 shows the radiative lifetime for the 2P system (averaged for levels $\nu'=0,1,2,3,4$) and 1N, taken from Table 19 of reference \cite{15Gilmore}, while the radiative lifetime measured for the $\upgamma$ system is displayed in Table IV of reference \cite{16Luque}.

\begin{table}
    \caption{Semi-empirical radiative transition parameters used in air radioluminescence modeling.}
    \centering
    \label{tab:Table1}
\begin{tabular}{|c||c|c|c|c|c|c|c|c|c|}
\hline
0 & 1 & 2 & 3 & 4 & 5 & 6 & 7 & 8 \\
BS & $\lambda_1$ & $\lambda_2$ & $\nu'$ & $\nu''$ & $q_{\nu'\rightarrow\nu''}$ & $A_{\nu'\rightarrow\nu''}$ & $B_{\nu'\rightarrow\nu''}$ & $\tau_{\nu'}$ \\
  & (nm) & (nm) &  &  &   & (s$^{-1}$) &   & (ns) \\
\hline\hline
\multirow{19}{*}{2P} & 337 & 337 &  & 0 & 4.54E-01 & 1.31E+07 & 4.86E-01 &  \multirow{19}{*}{38.4} \\
 & 357.4 & 357.6 &  & 1 & 3.27E-01 & 8.84E+06 & 3.28E-01 &  \\
 & 380 & 380.4 & \multirow{-3}{*}{0} & 2 & 1.45E-01 & 3.56E+06 & 1.32E-01 & \\
 \graycline
 & 315.6 & 315.8 & \multirow{5}{*}{1} & 0 & 3.92E-01 & 1.19E+07 & 4.46E-01 & \\
 & 334.9 & 333.8 &  & 1 & 2.26E-02 & 5.87E+05 & 2.20E-02 & \\
 & 353.5 & 353.6 &  & 2 & 2.05E-01 & 5.54E+06 & 2.08E-01 & \\
 & 375.2 & 375.4 &  & 3 & 1.98E-01 & 4.93E+06 & 1.85E-01 & \\
 & - & 399.7 &  & 4 & 1.10E-01 & 2.43E+06 & 9.11E-02 &  \\
 \graycline
 & 297.1 & 297.6 & \multirow{6}{*}{2} & 0 & 1.33E-01 & 3.97E+06 & 1.51E-01 & \\
 & 313.4 & 313.5 &  & 1 & 3.42E-01 & 1.01E+07 & 3.85E-01 & \\
 & - & 330.9 &  & 2 & 2.36E-02 & 7.99E+05 & 3.05E-02 & \\
 & - & 349.9 &  & 3 & 6.42E-02 & 1.71E+06 & 6.52E-02 &  \\
 & 370.4 & 370.9 &  & 4 & 1.61E-01 & 4.04E+06 & 1.54E-01 &  \\
 & 393.9 & 394.2 &  & 5 & 1.39E-01 & 3.14E+06 & 1.20E-01 &   \\
 \graycline
 & 281.7 & 281.8 & \multirow{4}{*}{3}& 0 & 2.02E-02 & 5.28E+05 & 2.06E-02 & \\
 & - & 296.1 &  & 1 & 2.53E-01 & 7.30E+06 & 2.85E-01 &  \\
 & 310.7 & 311.5 &  & 2 & 2.11E-01 & 5.94E+06 & 2.32E-01 & \\
 & 327.5 & 328.4 &  & 3 & 8.90E-02 & 2.85E+06 & 1.11E-01 & \\
 \graycline
 & 267.3 & 268.4 & 4 & 0 & 9.50E-04 & 1.38E+04 & 5.65E-04 & \\
 \hline
 1N & 391.1 & 391.2 & 0 & 0 & 6.63E-01 & 1.14E+07 & 7.10E-01 & 62.3 \\
\hline
\multirow{6}{*}{$\upgamma$} & 226.1 & 226.5 & \multirow{6}{*}{0} & 0 & 1.65E-01 & 9.26E+05 & 1.90E-01 & \multirow{6}{*}{206} \\
 & 236.2 & 236.6 &  & 1 & 2.62E-01 & 1.37E+06 & 2.81E-01 &  \\
 & 247.1 & 247.4 &  & 2 & 2.36E-01 & 1.15E+06 & 2.36E-01 &  \\
 & 258.7 & 259 &  & 3 & 1.60E-01 & 7.25E+05 & 1.49E-01 &  \\
 & 271.3 & 271.6 &  & 4 & 9.15E-02 & 3.86E+05 & 7.92E-02 &  \\
 & 285.1 & 285.3 &  & 5 & 4.65E-02 & 1.83E+05 & 3.75E-02 &  \\
    \hline

\end{tabular}

\end{table}

To use the parameters displayed in Table 1 as optical properties in the MPT, a \textit{photonEnergy} vector was created and its entries were sampled from the empirical data in column 1, by converting the wavelength into energy. For cases when a radioluminescence peak could not be resolved or the fitting results were unsatisfactory, the values from the second column were used. The first optical property is the scintillation component which refers to the emission probability. The input values describe the intensity of the spectral lines in the simulated radioluminescence spectrum. To quantify the probability that a molecule undergoes a vibronic transition $\nu'\rightarrow\nu''$, as well as shape the radioluminescence spectrum according to the relative emission probabilities, the product of the Franck-Condon coefficients (column 5) and branching ratios (column 7) was used. Three different scintillation components were created and registered in the MPT. These are the \textit{secondPositive}, \textit{firstNegative}, and \textit{gammaNO} vectors, which use as entries the values of the $q_{\nu'\rightarrow\nu''}\cdot B_{\nu'\rightarrow\nu''}$ product, corresponding to the 2P, 1N and $\upgamma$ regions, respectively. An optical property which does not depend on the energy vector is the scintillation yield, which was set to 19 photons per MeV \cite{10Sand}. To account for the quenching effects of the O$_2$ and H$_2$O molecules on the air radioluminescence spectrum \cite{1Luchkov}, the scintillation yield was separated into three relative yields, 0.65 (2P), 0.2 (1N), and 0.15 ($\upgamma$), which were attributed to their specific scintillation components. Lastly, the \textit{scintillation-time-constant} property was added for each region, using the values from column 8.

Within the OpNoviceDetectorDescription class, the geometry of the simulation was also defined. The scintillating medium was set to a cube with a side length of 50 meters containing air with a normal molecular composition, at 22°C and 1 atm pressure. At its centre, the geometry of the extended alpha source is described as a 30 mm $\times$ 100 mm Ag-foil with a thickness of 0.25 mm, on which an active layer of Am-241 is placed and covered with a 2 $\upmu$m Au-foil. The thickness of the Am-241 active area is 1 $\upmu$m and its width and length of 20 mm and 100 mm, respectively, are consistent with the Au layer \cite{1Luchkov}. The dimensions of the active area and the type of radionuclide used were set in the Am241.mac macro file. Similarly, the instructions for simulating the beam of alpha particles were written in alphabeam.mac. The simulated beam is characterized by a gaussian profile, and a diameter of 100 $\upmu$m.

The user classes OpNoviceSteppingAction, OpNoviceEventAction, and OpNoviceRunAction were modified to enable recording the simulated radioluminescence spectrum, the track length of the emitted alpha particles, and the deposited energy per alpha particle. The simulated values are registered in histograms and also be printed in data files. Lastly, the UV image of the alpha source is registered using a secondary radiation primitive scorer, set up to filter only optical photons. This method scores the number of photons traversing a defined surface and is implemented in the Am241.mac macro file.
\newpage
\section{Results and discussion}

The simulated radioluminescence spectrum was registered and scaled to the measurement results in \cite{4SandThes}, with respect to the 337 nm peak. Fig. 2 shows that the simulated spectrum is in good agreement with the experimental one in terms of wavelength and relative intensity of each peak. Although the relative yield used for simulating the UV-C component (0.15) leads to overestimating the scintillation yield in this region, it was the minimum value which could be used. Further lowering the relative yield led to no production of UV-C photons. However, without using relative yields, the simulation cannot properly account for the quenching effects of oxygen and water molecules. The deviations in the relative intensity of the simulated spectral lines compared to the measured peak intensities could be attributed to various factors. These include the absence of the quenching mechanisms for the N$_2$ and NO molecules in the physics models used, and the averaging of lifetimes corresponding to the $\nu'$ vibrational levels of the 2P system.

\begin{figure}[h!bt]
    \centering
    \includegraphics[width=0.8\linewidth]{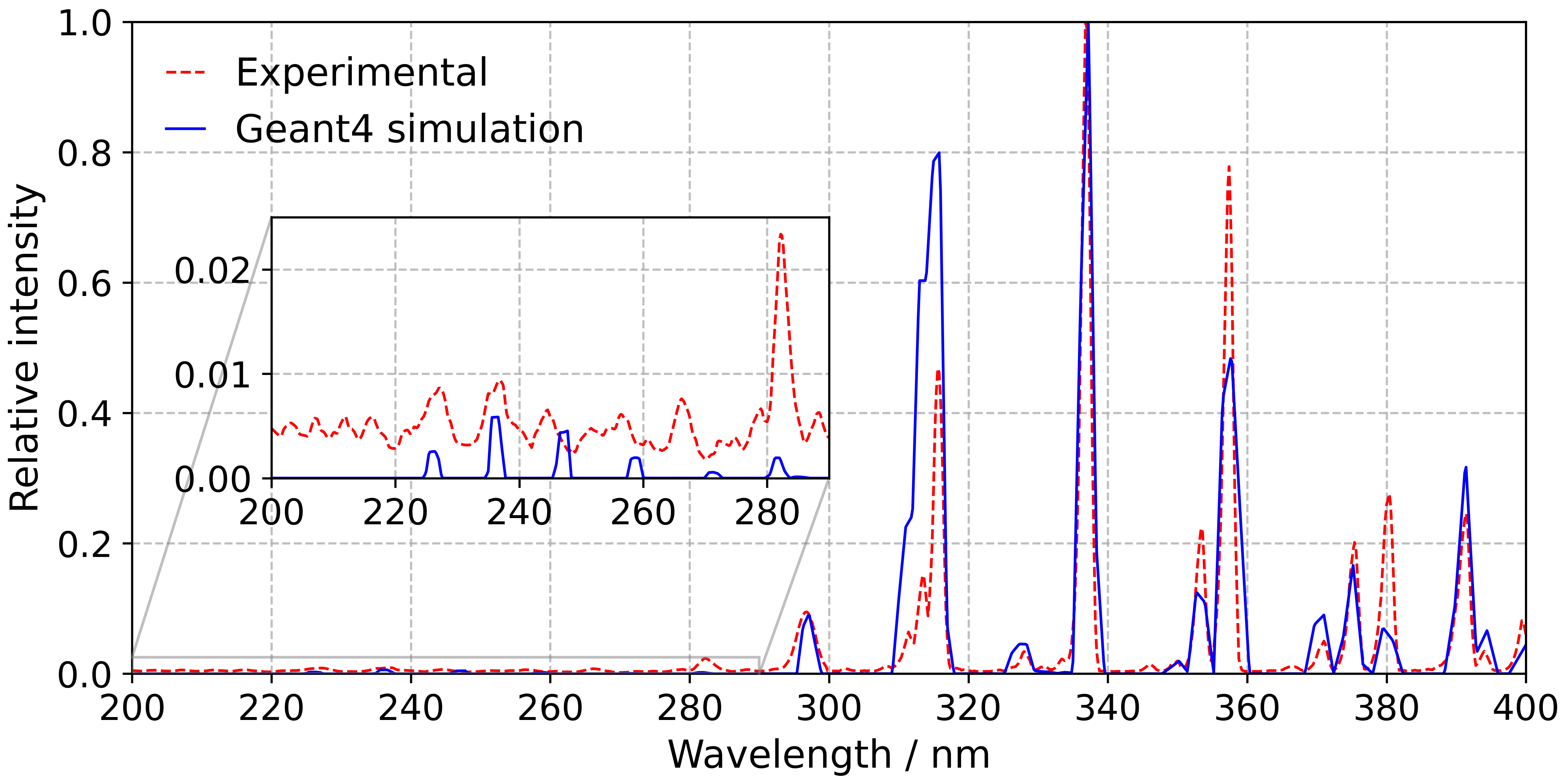}
    \caption{The radioluminescence emission spectrum of air spanning the UV region. The simulated spectrum is displayed in blue for a simulation of $10^6$ decays of the Am-241 source. The experimental spectrum measured in \cite{4SandThes} is presented in red for comparison. }
    \label{pic2}
\end{figure}

The simulation code was used to generate the radioluminescence images of the extended Am-241 source and the beams of alpha particles accelerated at 5.1 MeV and 8.3 MeV, depicted in Fig. 3. The track length of alpha particles was also computed. The results show that for a pure Am-241 source, the simulated mean track length is 41.7 ± 1.7 mm, which is lowered to 23.6 ± 7.6 mm in the case of the extended Am-241 source, due to an increased energy loss as alpha particles pass through the gold cover. The ultraviolet image of the Am-241 source shows that the boundaries of radioluminescence production include the length and width of the active area, as well as the mean track length of alpha particles.
In the case of alpha particles accelerated at 5.1 MeV, the computed mean track length of 37.67 ± 1.19 mm is consistent with the ultraviolet image of the Bragg peak illustrated in Fig. 3. As expected, by increasing the energy of the accelerated alpha particles to 8.3 MeV, the extent of the Bragg peak expands to approximately 80 mm. Therefore, the simulation predicts well both the length of the radioluminescence image and the location where the alpha particle energy loss is greatest (the Bragg peak).

The application effectively models the radioluminescence production in the proximity of alpha sources placed in air, allowing it to be used for optimization of radioluminescence detection technologies. Fig. 3 indicates that the application could be a useful tool in characterizing the mapping and imaging capabilities of radioluminescence setups. By coupling this method with ray-tracing simulations, and detector modeling, the optimization and development of radioluminescence setups could be further achieved.

\begin{figure}[h!bt]
    \centering
    \includegraphics[width=0.8\linewidth]{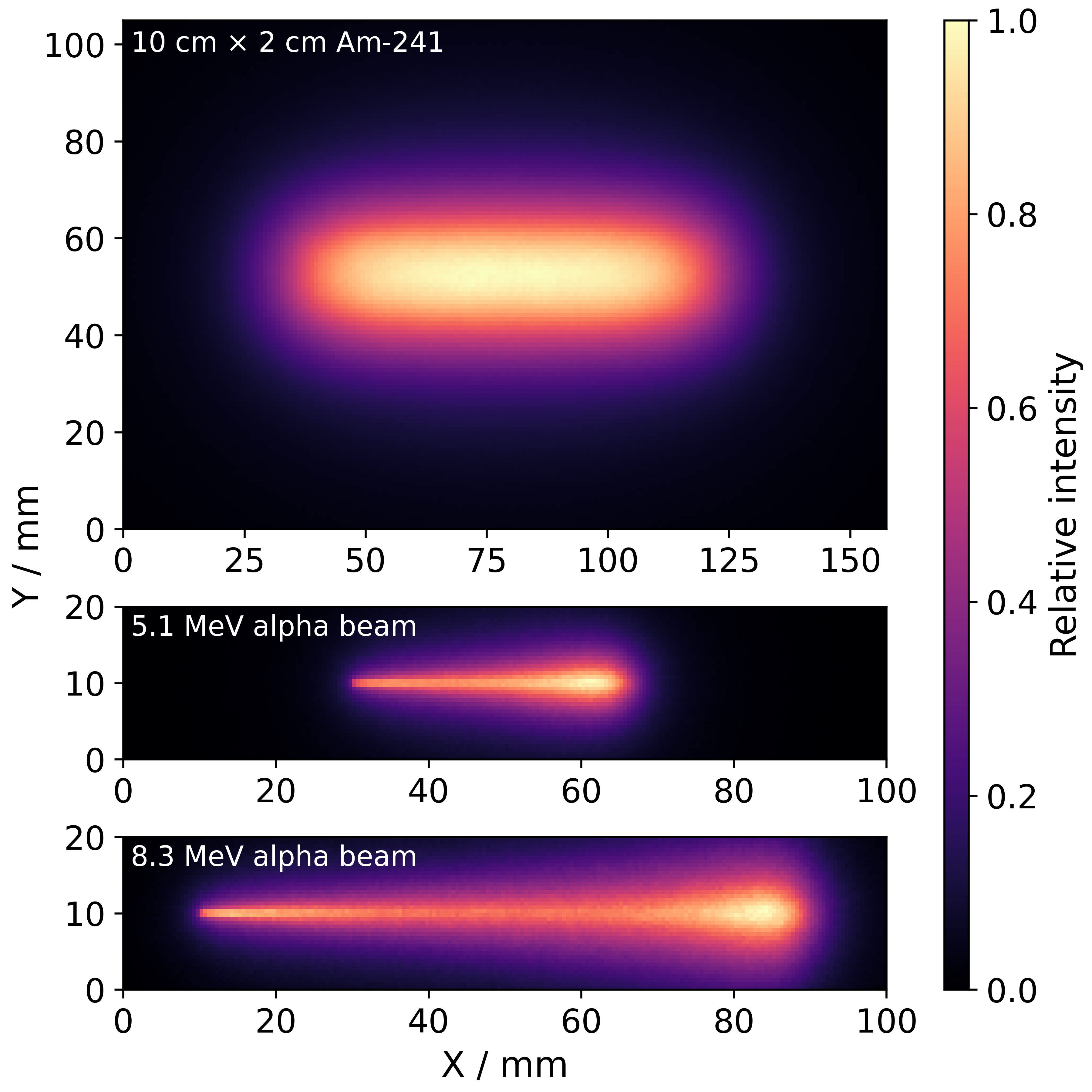}
    \caption{The ultraviolet images of the simulated Am-241 source (top) and the simulated beam of alpha particles accelerated at 5.1 MeV (middle) and 8.3 MeV (bottom). }
    \label{pic3}
\end{figure}
\newpage
\section{conclusion}
A Geant4 \cite{17geant} application was developed based on the OpNovice example from the toolkit package. It demonstrates an accessible method to simulate radioluminescence production using presented input parameters - the radiative transition parameters specific to N$_2$ and NO molecules, responsible for air luminescence. In the context of the RemoteALPHA project, the simulation was designed to model the radioluminescence effect by employing two types of alpha sources, an extended Am-241 sample and accelerated alpha particles. The application was demonstarted to compute the spectrum of emitted optical photos, as well as the ultraviolet image of the alpha sources, indicating that it can serve as a tool in optimizing remote alpha detection systems. To support future research in this field, the application is available to readers at \cite{18github}.

\begin{acknowledgement}
The authors wish to express their gratitude to the European Association of National Metrology Institutes (EURAMET) and the Physikalisch-Technische Bundesanstalt (PTB) for their support in the participation of Claudia Olaru in the research activities related to the EMPIR Researcher Mobility Grant (RMG). They also extend their appreciation to Maksym Luchkov, Dr. Faton Krasniqi, Dr. Volker Dangendorf, and Dr. Ulrich Giesen of PTB for their assistance during Claudia Olaru's RMG visit at PTB and their support with experiments at the PTB Ion Accelerator Facility. Additionally, the authors are thankful to Prof. Dr. Ionel Lazanu of the University of Bucharest, Faculty of Physics, for his unwavering support.

The project 19ENV02 RemoteALPHA has received funding from the EMPIR programme co-financed by the Participating States and from the European Union’s Horizon 2020 research and innovation programme. 19ENV02 RemoteALPHA denotes the EMPIR project reference. 

This work was partly funded by the Romanian Ministry of Research, Innovation and Digitalization, from the Core Project PN: 23 21 02 03. 

\end{acknowledgement}

\end{document}